\documentclass{aastex63}

\usepackage{amsmath}
\newcommand{\Pram}{$P_\text{ram}$}

\usepackage{xcolor}
\newcommand{\COMON}{\begin{color}{blue}}
\newcommand{\COMOFF}{\end{color}}

\submitjournal{ApJ}

\shorttitle{Collapse of Mercury's Dayside Magnetosphere}
\shortauthors{Winslow et al.}

\begin{document}

\title{Observations of Extreme ICME Ram Pressure Compressing Mercury's Dayside Magnetosphere to the Surface}

\correspondingauthor{Reka M. Winslow}
\email{reka.winslow@unh.edu}

\author{Reka M. Winslow}
\affiliation{Institute for the Study of Earth, Ocean, and Space, University of New Hampshire, Durham, NH, USA}

\author{No\'e Lugaz}
\affiliation{Institute for the Study of Earth, Ocean, and Space, University of New Hampshire, Durham, NH, USA}

\author{Lydia Philpott}
\affiliation{Department of Earth, Ocean and Atmospheric Sciences, The University of British Columbia, Vancouver, BC, Canada}

\author{Charles J. Farrugia}
\affiliation{Institute for the Study of Earth, Ocean, and Space, University of New Hampshire, Durham, NH, USA}

\author{Catherine L. Johnson}
\affiliation{Department of Earth, Ocean and Atmospheric Sciences, The University of British Columbia, Vancouver, BC, Canada}
\affiliation{Planetary Science Institute, Tucson, AZ, USA}

\author{Brian J. Anderson}
\affiliation{The Johns Hopkins University Applied Physics Laboratory, Laurel, MD, USA}

\author{Carol S. Paty}
\affiliation{Department of Earth Sciences, University of Oregon, Eugene, OR, USA}

\author{Nathan A. Schwadron}
\affiliation{Institute for the Study of Earth, Ocean, and Space, University of New Hampshire, Durham, NH, USA}

\author{Manar Al Asad}
\affiliation{Department of Earth, Ocean and Atmospheric Sciences, The University of British Columbia, Vancouver, BC, Canada}



\begin{abstract}
 Mercury's magnetosphere is known to be affected by the enhanced ram pressure and magnetic fields inside interplanetary coronal mass ejections (ICMEs). Here we report detailed observations of an ICME compressing Mercury's dayside magnetosphere to the surface. A fast CME launched from the Sun on November 29 2013 impacted first MESSENGER, which was orbiting Mercury, on November 30 and later STEREO-A near 1 AU on December 1. Following the ICME impact, MESSENGER remained in the solar wind as the spacecraft traveled inwards and northwards towards Mercury's surface until it reached and passed its closest approach to the planet (at 371 km altitude) without crossing into the magnetosphere. The magnetospheric crossing finally occurred 1 minute before reaching the planet's nightside at 400 km altitude and 84$^\circ$N latitude, indicating the lack of dayside magnetosphere on this orbit. In addition, the peak magnetic field measured by MESSENGER at this time was 40\% above the values measured in the orbits just prior to and after the ICME, a consequence of the magnetospheric compression. Using both a proxy method at Mercury and measurements at STEREO-A, we show that the extremely high ram pressure associated with this ICME was more than high enough to collapse Mercury's weak magnetosphere. As a consequence, the ICME plasma likely interacted  with Mercury's surface, evidenced by enhanced sodium ions in the exosphere. The collapse of Mercury's dayside magnetosphere has important implications for the habitability of close-in exoplanets around M dwarf stars, as such events may significantly contribute to planetary atmospheric loss in these systems. 

\end{abstract}

\keywords{}


\section{Introduction} 

Mercury is the only planet in the inner solar system, other than Earth, that possesses a dynamo-generated global, albeit weak, magnetic field \citep[]{Ness1975,Anderson2011}. Mercury's proximity to the Sun, with a heliocentric distance between 0.31 and 0.47 au, results in more extreme solar wind and interplanetary conditions than at any other planet in our solar system. As a consequence of this weak internal magnetic field and strong upstream conditions, the magnetosphere of Mercury is highly dynamic \citep[e.g.,][]{SlavinHolzer1979,Glassmeier2007}. This is especially true during times of interplanetary coronal mass ejections (ICMEs), which travel at high speeds compared to the solar wind (up to 3000~km\,s$^{-1}$) and carry mass and magnetic field from the Sun \citep[]{Manchester2017}. It has long been hypothesized \citep[e.g.,][]{SlavinHolzer1979, Kabin2000,Slavin2010} that the magnetic flux on Mercury's dayside may be completely eroded or compressed below the surface during such extreme conditions, and that erosion via reconnection is likely the dominant physical process causing this.

The MErcury Surface, Space ENvironment, GEochemistry, and Ranging \citep[MESSENGER][]{Solomon2007} spacecraft orbited Mercury between March 2011 and April 2015. Its highly eccentric orbit around Mercury had periapsis altitudes between 200-500 km and apoapsis altitudes of 10,000-12,000 km. This allowed the spacecraft to spend a significant fraction of each orbit both in the solar wind and in the magnetosphere. The years MESSENGER orbited Mercury corresponded to the maximum phase of solar cycle 24. As such, MESSENGER was able to directly measure the effects of ICMEs on Mercury's magnetosphere. Over the duration of the mission, 69 ICMEs were detected by MESSENGER \citep[]{Winslow2015,Winslow2017}.

Mercury's weak internal field \citep{Ness1975,Anderson2011,Johnson2012}, described by a dipole with a moment of 190 nT $R_M^3$ (where $R_M$ is Mercury's radius), offset 0.2 $R_M$ northward from the geographic equator \citep[]{Anderson2011,Anderson2012,Johnson2012,Winslow2014}, does not present a large obstacle to the solar wind. During average solar wind conditions \citep[]{Winslow2013}, the magnetopause subsolar stand-off distance, $R_{SS}$, is 1.45 $R_M$ from the planet's center, and the bow shock subsolar stand-off distance is 1.96 $R_M$, about 5-6 times closer to the planet than those of Earth. These distances are largely controlled by solar wind parameters \citep{Winslow2013} and are most strongly affected during ICMEs when conditions are extreme \citep[]{Slavin2014,Winslow2017}. In \citet{Winslow2017}, we described the average effect of CMEs on Mercury's magnetosphere based on 113 MESSENGER orbits affected by ICMEs. We concluded that due to enhanced dynamic pressure during ICMEs, the average subsolar location of the magnetopause was 15\% closer to the planet. Based on the typical shape of Mercury's boundaries, these measurements indicate that the magnetopause was likely at the planet's surface during $\sim$30$\%$ of ICME-affected MESSENGER orbits \citep[]{Winslow2017}. Comparable estimates based on all orbits (not only those affected by ICMEs) suggest this occurs $\sim 1.5$--$4\%$ of the time \citep[]{Johnson2016}. We also determined that the solar wind had an average ram pressure of 86 nPa  during the ICME-affected orbits for which the magnetopause may have reached the surface, 6 times larger than typical conditions. 


In this paper, we examine in detail unambiguous evidence of the collapse of Mercury's dayside magnetosphere due to
the highest ever ram pressure values estimated at Mercury that were associated with an ICME in late November 2013. The ICME affected two consecutive orbits of MESSENGER around Mercury. Because the magnetospheric perturbation was strongest during the first orbit, we analyzed this orbit (orbit 2577) in detail (Figure 1a). In a paper published independently of this work, \cite{Slavin2019} also analyzed the effect of this ICME on Mercury's magnetosphere. However, their work focused on the second ICME-affected orbit which displayed less extreme solar wind conditions and magnetospheric response than in the first ICME-affected orbit analyzed in this study (Figure 1a). \cite{Slavin2019} emphasized flux transfer events \citep[FTEs][]{RussellElphic1978,Slavin2012}, whereas we focus on demonstrating unambiguously, the compression of Mercury's dayside magnetosphere to the surface.  The organization of this paper is as follows: in Section~\ref{sec:ram}, we discuss the upstream solar wind measurements, both by MESSENGER but also at STEREO-A which was radially aligned with MESSENGER and provides a more comprehensive view of this event. In Section~\ref{sec:MP}, we present direct evidence that the dayside magnetopause of Mercury was compressed down to the planetary surface. In Section~\ref{sec:enhancements}, we discuss the magnetospheric conditions, both in terms of the enhanced peak fields as well as the conditions of the magnetotail. We discuss our results and conclude in Section~\ref{sec:conclusion}.

\begin{figure}
\begin{center}
\includegraphics[scale=0.6]{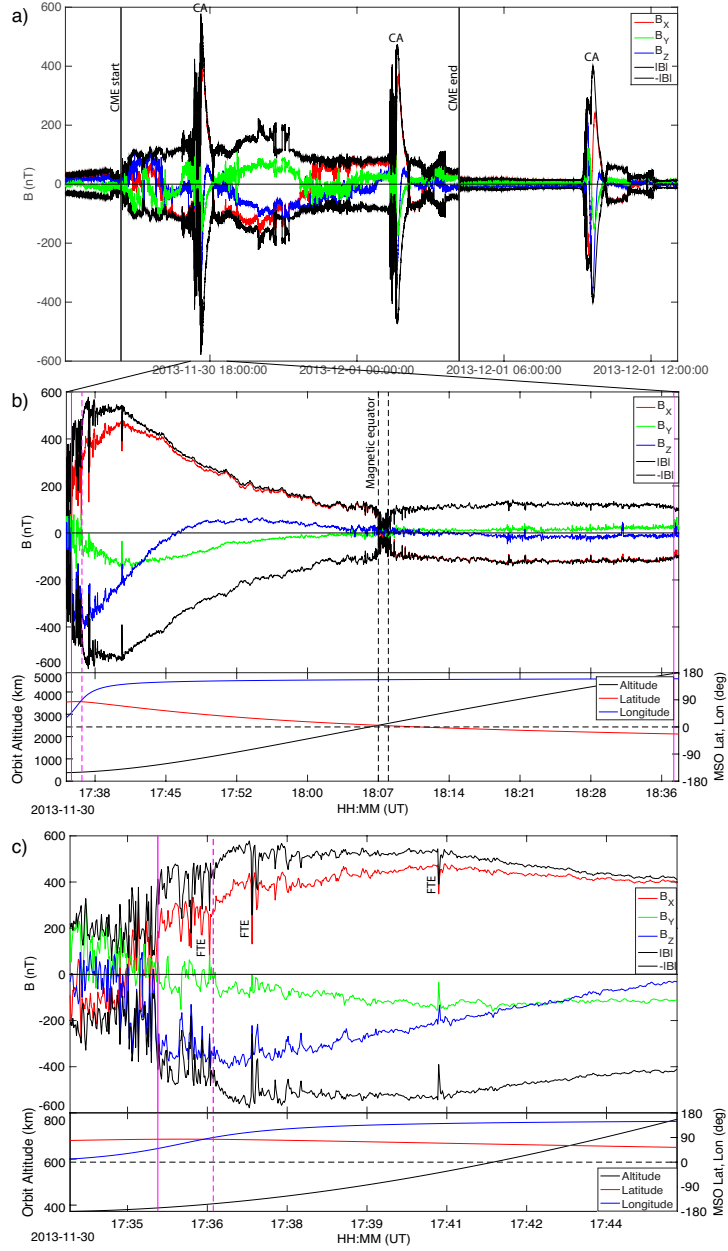} 
\caption{{\bf{MESSENGER magnetic field observations during the ICME impact at Mercury.}} a) Data from three orbits of MESSENGER around Mercury, with the ICME arrival and end marked by black vertical lines and the spacecraft closest approach is marked by CA. b) The magnetospheric pass on the first orbit from panel a). Vertical solid magenta lines indicate inbound and outbound magnetopause crossings, while dashed magenta line indicates 90$^{\circ}$ longitude, i.e. beginning of the night-side transit. Vertical dashed black lines mark the magnetic equator crossing. c) The spacecraft closest approach region of the magnetospheric pass shown in b). Magenta lines are same as in  
b).} 
\end{center}
\end{figure}

\section{Solar wind ram pressure}\label{sec:ram}
The ICME was launched from the Sun on November 29 2013. At that time, MESSENGER was at a distance of 0.40 au from the Sun and $\sim$ 120$^\circ$ west of the Sun-Earth line and the STEREO-A spacecraft \citep[]{Kaiser2008} was at a distance of 0.96 au and $\sim$ 150$^\circ$ west of the Sun-Earth line. The ICME was moderately fast, with a speed in the corona of about 700-900~km\,s$^{-1}$. The detailed analysis of the initial phase of the ICME propagation, including its direction, kinematics and the formation of a CME-driven shock has been published in \citet{Schmidt2016}. The CME was found to be launched about 135$^\circ$ from the Sun-Earth line, so almost exactly halfway between Mercury and STEREO-A directions. The ICME reached Mercury at 14:25:35 UT on November 30 2013, and was observed {\it in situ} by the Magnetometer \citep{Anderson2007} onboard MESSENGER (Figure 1a) to affect two consecutive MESSENGER orbits around Mercury (orbits 2577 and 2578). During orbit 2577, the periapsis altitude of MESSENGER was 371 km while the apoapsis altitude was 10,302 km, and MESSENGER was in an 8-hour orbit.

The magnetic field direction in the magnetic cloud was predominantly southward (-$B_Z$), however, it had a northward ($B_Z$) component for $\sim$30~mins prior to MESSENGER crossing into Mercury's magnetosphere on this first ICME-affected orbit. In discussion of MESSENGER magnetic field data we use MSO coordinates, where $X_{MSO}$ is positive sunward, $Z_{MSO}$ is positive northward, $Y_{MSO}$ is positive duskward and completes the right-handed system, and the origin is at the center of the planet.

Due to the lack of direct solar wind observations by MESSENGER \citep{Andrews2007}, we first used proxy methods to estimate the relevant ICME parameters at this time. Using the modified Newtonian approximation described in equation (4) of \citet{Winslow2017}, we estimated the ram pressure, \Pram, to be $385\pm177$ nPa from the inbound magnetopause crossing on orbit 2577. The error bar reflects the uncertainty in the angle between the magnetopause normal and the upstream solar wind direction, as well as the uncertainty in the fraction of the solar wind ram pressure that gets transferred into the magnetosheath. The magnetopause crossing position was identified following the approach outlined in \citet{Winslow2013}.
 In comparison, the average \Pram\ at Mercury is 14~nPa, nearly thirty times lower. From the ICME launch time at the Sun \citep{Winslow2015} and the arrival time at Mercury, we calculated the Sun-to-Mercury transit speed to be $\sim$800~km~s$^{-1}$, which together with the \Pram\ estimate yields a plasma number density of $\sim$360~cm$^{-3}$, compared to $\sim$50~cm$^{-3}$ on average. 

As the STEREO-A spacecraft was in near-longitudinal alignment with MESSENGER at this time, this same ICME was also observed by STEREO-A near 1 au. 
The STEREO spacecraft have a full suite of plasma instruments taking continuous measurements of the solar wind \citep{Galvin2008}, which can be scaled to Mercury's location to provide an independent estimate of the ram pressure during this ICME. Assuming an alpha to proton ratio (He/H) in the plasma of 5$\%$, which is typical of the solar wind, the maximum \Pram\ of 38 nPa at STEREO A (Figure 2) yields 234 nPa at MESSENGER, with a scaling of $1/r^2$ with heliocentric distance. However, He/H ratios as high as 33$\%$ were observed by STEREO-A during this ICME (private communication, A.B. Galvin, 2019), yielding a \Pram\ of 63 nPa at 1 au and 388 nPa at Mercury (at 0.4 au). These scaled STEREO A observations are in agreement within uncertainty with the \Pram\ proxy value we obtained from the Newtonian approximation at MESSENGER, fully supporting the estimate of a very high \Pram\ during this ICME. The \Pram\ measured near 1 au corresponds to one of the highest 0.01\% ram pressure values measured at L1 during the 4 years of the MESSENGER mission. Using spacecraft measurements at Earth's L1 during these 4 years, \Pram  reached as high as 35 nPa for only 8 distinct 5-minute periods out of ~395,000 such periods. All this supports the fact that this was an extremely high ram pressure event both at Mercury and 1 au. 

\begin{figure}
\begin{center}
\includegraphics[scale=0.4]{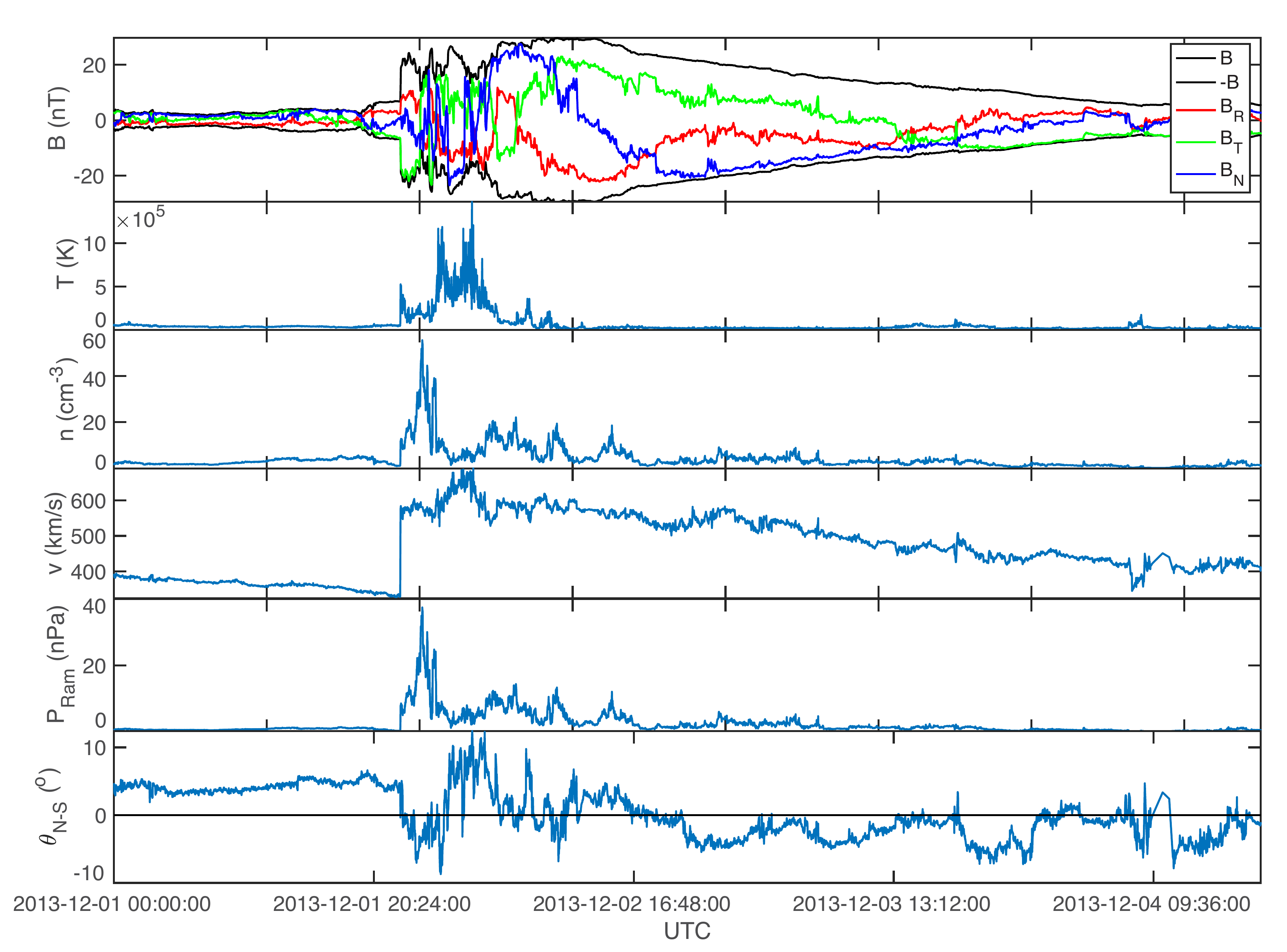} 
\caption{{\bf{STEREO A measurements of the same ICME near 1 au.}} From top to bottom, the panels show the magnetic field magnitude and components, the plasma temperature, plasma density, plasma speed, the ram pressure of the plasma, as well as the $\theta_{north-south}$ angle. The magnetic field vector components are shown in heliospheric radial-tangential-normal (RTN) coordinates. The \Pram\ value shown assumes an alpha to proton ratio of 5$\%$ in the plasma, which is the lower limit during this ICME event. } 
\end{center}
\end{figure}

\section{Magnetopause compression to the surface}\label{sec:MP}

The most remarkable effect of this ICME on Mercury was the complete compression of the dayside magnetopause to the surface (Figures 1 and 3) resulting from the very high \Pram. During noon-midnight orbits, MESSENGER regularly spent $\sim$20 minutes sampling Mercury's low-latitude dayside magnetosphere, usually entering the magnetosphere at latitudes close to the equator. Even though MESSENGER was in a noon-midnight orbit during this ICME, the location of the dayside magnetopause crossing into the magnetosphere occurred at a high latitude of 84$^{\circ}$ N, a longitude of 50$^{\circ}$, and an altitude of only $\sim$400 km (Figures 1b,c and Figure 3). Prior to this crossing, MESSENGER was at altitudes less than 400 km at more southern latitudes and the magnetosphere was not encountered (spacecraft closest approach occurred at an altitude of 371 km and a latitude of 76.7$^{\circ}$ N). Less than 1 minute after crossing the magnetopause, MESSENGER crossed the dawn-dusk terminator and entered the nightside. Thus unlike typical noon-midnight orbits, MESSENGER spent less than 1 minute in the dayside after crossing the magnetopause, which explains the lack of a cusp crossing. No dayside northward magnetic field direction was observed in the magnetosphere (Figure 1b,c and Figure 3). The field lines that MESSENGER sampled immediately after entering the magnetosphere were southward and sunward pointing, consistent with field lines in the high northern hemisphere close to and threading the night-side. We also note that there was a very thin inbound magnetopause layer on this orbit that lasted $<2$ mins, indicating quasi-steady external forcing conditions at this time.

\begin{figure}
\begin{center}
\includegraphics[scale=0.4]{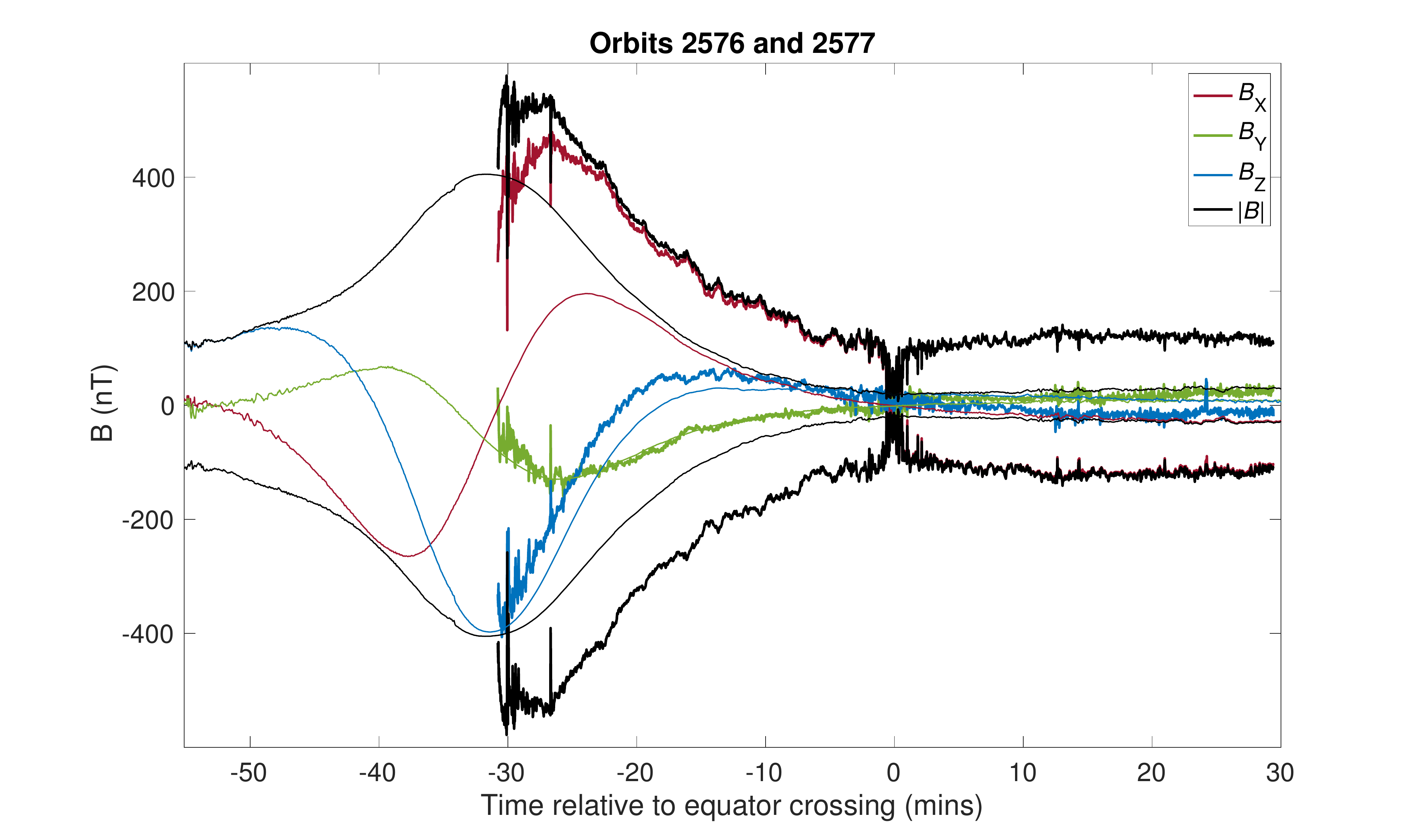} 
\caption{{\bf{Comparison of MESSENGER's magnetospheric pass on the pre-ICME orbit with the first ICME affected orbit.}} Magnetic field data versus time relative to the magnetic equator crossing times for orbits 2576 (thin lines) and 2577 (thick lines). Equator crossing times were at UTC 10:07:02 and UTC 18:07:49 for orbits 2576 and 2577, respectively.} 
\end{center}
\end{figure}

In Figure~3, MESSENGER orbits 2576 (the orbit just prior to the ICME arrival) and 2577 (the first orbit affected by the ICME in question) were aligned in time on their equator crossing times to compare the magnetic field data before and after the ICME arrival. The increase in field strength in the nightside magnetosphere after the ICME arrival is evident from orbit 2576 to 2577, with peak field strengths of $\sim$405 nT and $\sim$575 nT on orbits 2576 and 2577, respectively.  Furthermore, the almost complete absence of the dayside field on orbit 2577 compared to orbit 2576 is striking: the dayside northern hemisphere dipole field on orbit 2576 is seen via downward pointing (negative $B_Z$), anti-sunward (negative $B_X$) field, that was absent on orbit 2577.

The established shape and location of Mercury's magnetopause and bow shock determined empirically provide further confirmation of the extreme compression. 
Similarly to \citet{Winslow2013,Winslow2017}, we fit empirical models to the magnetopause and bow shock crossing points to characterize the magnetospheric boundaries during orbit 2577. For the bow shock, we used a conic section given by 

\begin{equation}
\sqrt{(X-X_0)^2+\rho^2} = \frac{p\epsilon}{1+\epsilon\cdot \cos\theta},
\end{equation}

\noindent where $X_0$ is the focus point, $\epsilon$ is the eccentricity, $p$ is the focal parameter, $\rho = \sqrt{Y^2 + (Z-Z_{d})^2}$ and $\theta=\tan^{-1}(\frac{\rho}{X})$. Here $Z_{d} = 484\,\mathrm{km}$ is the offset of the dipole from the planetary equator. For the magnetopause, we used the empirical model by Shue et al.\citep{Shue1997}

\begin{equation}
R = \sqrt{X^2+\rho^2} = R_{ss}\left(\frac{2}{1+\cos\theta}\right)^{\alpha},
\end{equation}

\noindent where $R$ is the distance from the dipole center and $\alpha$ is the flaring parameter that governs how open the magnetotail is. The average magnetopause shape at Mercury \citep{Winslow2013} has $\alpha = 0.5$, and $\alpha > 0.5$ implies an open magnetosphere while $\alpha < 0.5$ implies a closed magnetosphere.

\begin{figure}
\begin{center}
\includegraphics[scale=0.5]{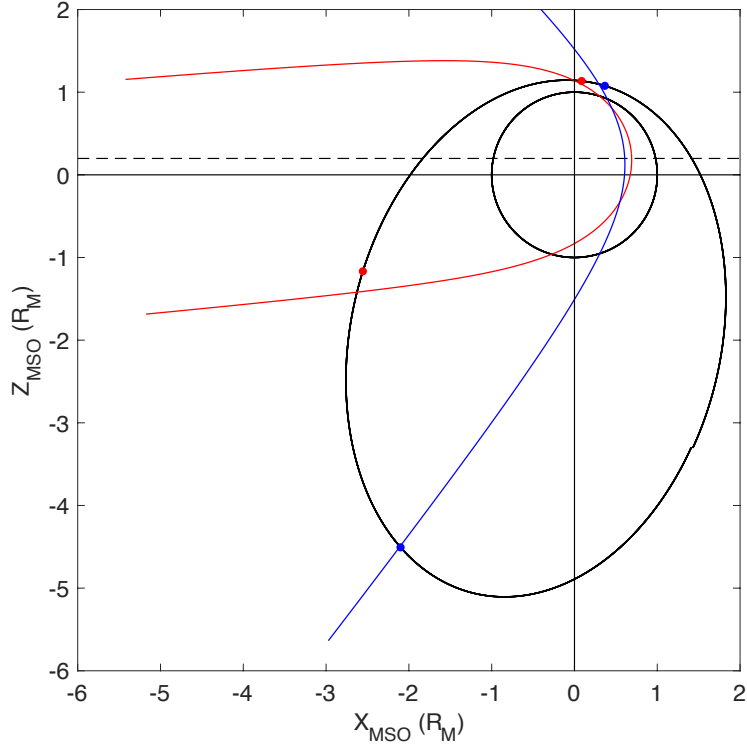}
\caption{{\bf{MESSENGER's orbit and magnetospheric boundaries shortly after the ICME arrival.}} MESSENGER's noon-midnight orbit around Mercury right after the ICME arrival is shown in black in the $Z_{MSO}-X_{MSO}$ plane.  Best-fit magnetopause (red) and bow shock (blue) boundaries are shown, along with the tilt caused by the non-radial solar wind flow. The magnetopause and bow shock boundary crossings are marked by red and blue dots, respectively. Note that the crossing locations do not occur exactly at $Y_{MSO} = 0$, and thus when viewed in the $Z_{MSO}-X_{MSO}$ plane, they do not lie directly on the profile curves of the best-fit boundaries. The dashed line represents the dipole offset of $Z_d$ = 484 km from the planetary equator.} 
\end{center}
\end{figure}

The \citet{Shue1997} magnetopause model has been shown to fit Mercury's average magnetopause shape well, under both nominal and extreme solar wind conditions \citep{Winslow2013,Winslow2017}. We use this axially symmetric model to illustrate the possible magnetospheric shape of Mercury at this time, and we note that a 3 dimensional model specifically developed for Mercury \citep{Zhong2015} does exist, but contains additional parameters that cannot be constrained by the data available here. 

We used the above models to fit our ICME-affected bow shock and magnetopause crossing locations, corrected for solar wind aberration, using a grid search method that minimized the root mean square (RMS) residual of the perpendicular distance of the observed boundary crossing from the model boundary. Although not the focus of the paper, these fits indicate that the bow shock was also substantially compressed on orbit 2577 (Figure 4), having an estimated subsolar stand-off distance of $0.61~R_M$ with best-fit model parameters of $X_0=0$, $\epsilon=1.46$, and $p=1.02~R_M$. This fit result implies a bow shock with a stand-off distance that is below the surface and is smaller than that of the magnetopause. Even in the absence of a dayside magnetopause, one still expects a bow shock to form above the planetary surface due to the obstacle of the planet itself; the bow shock is needed to deflect the plasma around it. This does suggest that the model breaks down for such extreme solar wind conditions, although the fit is likely correct for the part of the orbit where the crossing points actually are (i.e. where the fit is done).

For the inbound and outbound aberrated magnetopause crossings on this orbit, the best-fit Shue et al. model yields an $R_{ss}$ of 0.69 $R_{M}$, i.e. a value that is well below the surface (Figure 4; see Section 4.2.1 for tilting of the magnetotail). It is conceivable that a different magnetopause shape with a very large cusp (such as is possible in the \citet{Zhong2015} model) might be below MESSENGER’s orbit but still be above the surface in some of the northern hemisphere. Without a probe on the surface, such a scenario cannot be ruled out. However, combining different lines of evidence, such as the measured bow shock (see Figure 4) and magnetopause locations, the very large inferred \Pram, and the fact that the magnetopause was not observed at any latitude below 84$^{\circ}$\,N even though the lowest altitude reached by MESSENGER during this orbit occurred at 76.7$^{\circ}$\,N, the evidence clearly supports the complete absence of a dayside magnetosphere in this case (Figure 5).

\begin{figure}
\begin{center}
\includegraphics[scale=0.4]{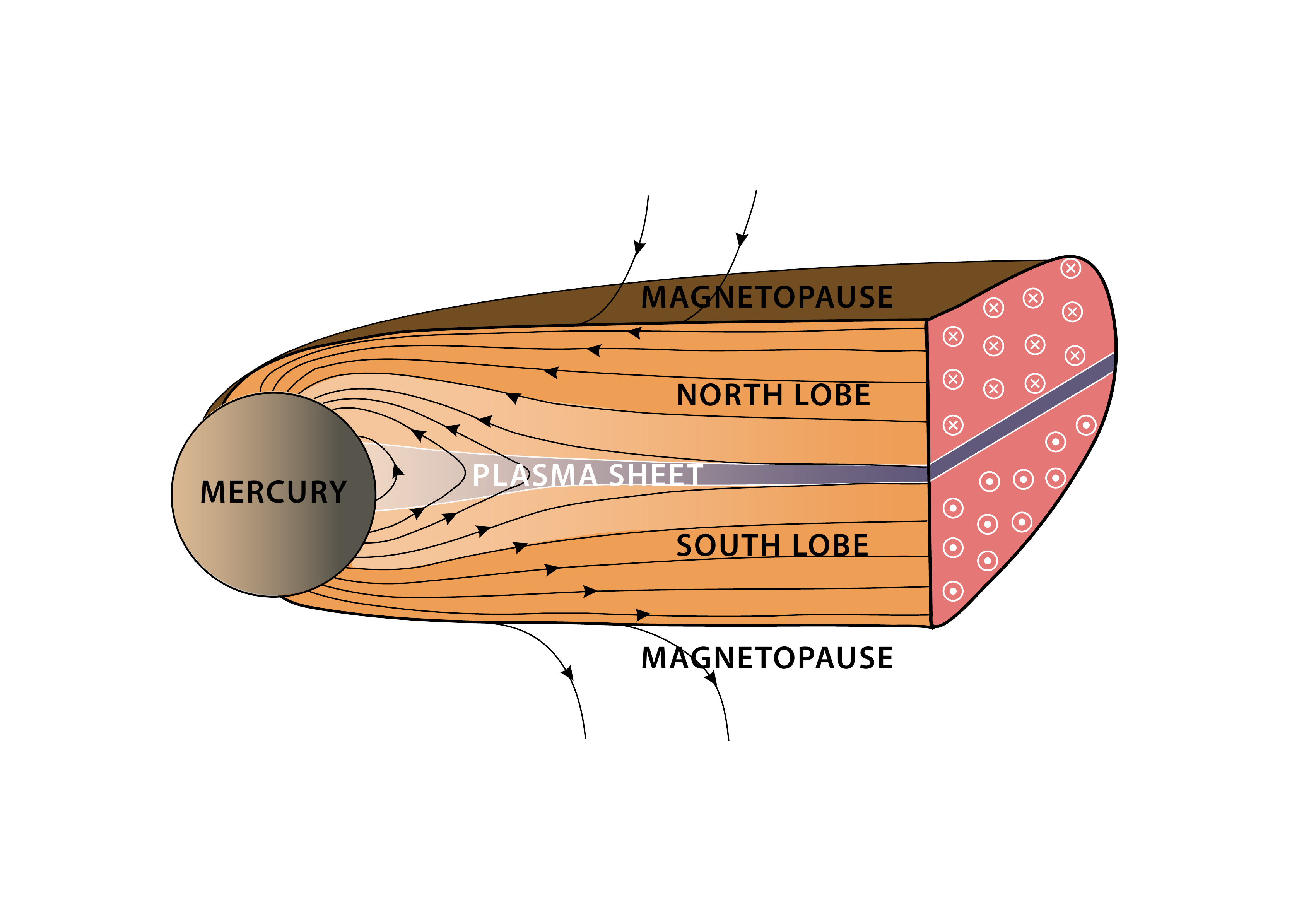} 
\caption{{\bf{Schematic view of Mercury's magnetosphere, shown in the frame of reference of the solar wind flow, during the first ICME affected orbit.}} The ICME compressed the dayside magnetosphere to the surface, leaving the region directly exposed to the solar wind. A larger surface area is exposed in the southern hemisphere than in the north due to Mercury's northward-offset dipole.} 
\end{center}
\end{figure}

For the model fits to the boundary locations, we used the instantaneous boundary crossings nearest to the magnetosphere for both the inbound and outbound crossings, and fit the inbound and outbound points with one model. Fitting the inbound and outbound magnetopause crossings with one magnetopause model assumes that the solar wind conditions do not change substantially in the 1 hr between the inbound and outbound observations. This assumption is not unreasonable for an ICME passage, during which the solar wind plasma and magnetic field parameters are expected to vary more slowly than under average solar wind conditions.

From statistical analysis of the response of Mercury's magnetopause to solar wind ram pressure using MESSENGER observations, \citet{Winslow2013} estimated that a \Pram\ of 175\,nPa would be enough to collapse Mercury's magnetosphere to the surface. In the global MHD simulations conducted by \citet{Kabin2000}, a \Pram\ of 147\,nPa was found to accomplish this task, however, this model did not include induction effects from Mercury's core. \citet{Jia2019} recently included induction effects in MHD simulations of the Mercury-solar wind interaction. They found that in simulations without induction, 56 nPa was sufficient for the magnetopause to reach the southern hemisphere surface, while if induction was included, most of the magnetosphere was above the surface for the highest tested \Pram\ of 107\,nPa. This shows that the strong interplay between induction and compression has an important effect on the size of Mercury's magnetosphere. Overall, these studies predict compression to Mercury's surface at \Pram\ $\sim$ 150--175 nPa. Given the much higher \Pram\ of $385\pm177$ nPa estimated for the ICME in our study and supported by the STEREO-A scaled observations, it is not surprising that it collapsed Mercury's dayside magnetosphere. Also, using our estimated $\sim$385 nPa \Pram, we find that the pressure balance condition of the magnetopause \citep{Kallenrode2004}, which does not depend on any model assumption for the magnetopause shape, results in a magnetopause subsolar stand-off distance of 0.74 $R_M$.

We also calculate the surface area of the region directly exposed to the solar wind to be $\sim50\%$ of the total planetary surface. This estimate is obtained from where our best-fit Shue et al. magnetopause model (Figure 4) intersects the planetary surface in the northern and southern hemispheres, with the northward offset dipole taken into consideration. We note that this estimate is dependent on the magnetopause model used, and it may be an upper limit given that the Shue model does not account for any possible cusp indentations of the magnetopause.

\section{Magnetospheric Conditions}\label{sec:enhancements}

\subsection{Field Enhancements}
Another remarkable effect of this ICME on Mercury's magnetosphere was the $\sim$40$\%$ increase in the observed field strength at the spacecraft's closest-approach (represented by the peak in the field) compared to orbits prior to, and after the ICME (Figure 1a). We investigated whether this increased field strength could be attributed to an induction effect resulting from magnetospheric compression. Annual induction signals have been observed in MESSENGER data \citep{Johnson2016} and larger signals are expected during times of extreme compression \citep{Glassmeier2007}. For example, self-similar compression of the magnetosphere (uniform scaling in X, Y, and Z), corresponding to a change in subsolar distance from $1.6 R_M$ to $1 R_{M}$ could result in an increase in the internal dipole moment of up to $\sim$25$\%$, similar to the observed increase in the field strength. However, an induction effect would increase the magnitude of all three components close to the planet, which was not observed. Figure 3 shows that the increase in the field was almost entirely due to an increase in $B_X$, which persisted throughout the orbit. In contrast, the $B_Y$ component was almost identical on both orbit 2577 and the orbit prior to the ICME arrival, and $B_Z$ decreases slightly close to the planet. In the tail, the field magnitude was enhanced by a factor of $\sim$4 on orbit 2577 compared to 2576, suggesting that very different tail conditions (specifically a very strong magnetotail current, see Section 4.2.2 below), not induction, dominate the observed signal.

\subsection{Magnetotail Conditions}

In this section we explore Mercury's magnetotail conditions during this first ICME-affected orbit. Despite the enhancement in the magnetotail field during this event, we found no evidence for flaring of the tail (Section 4.2.1 below). This, together with the calculated tail flux of $2.73\pm0.86$~MWb (see Section 4.2.2 below), consistent with the average value of $2.6\pm0.6$~MWb observed at Mercury \citep{Johnson2012}, shows that the increase in the magnetotail field strength was not caused by flux-loading of the tail. The absence of flaring implies that Dungey cycle magnetospheric convection did not substantially enhance the tail field, which is supported by the fact that there are only a few FTEs. These FTEs are observed near the inbound magnetopause between latitudes of 65 to 90 deg, indicating some high-latitude reconnection. Low-latitude reconnection is ruled out because of the lack of dayside magnetosphere. However, observations by the Fast Imaging Plasma Spectrometer (FIPS) \citep{Andrews2007} onboard MESSENGER indicate that plasma was present in the southern tail lobe (Figure 7), likely convected into the magnetosphere through the plasma mantle or via lobe convection driven by high-latitude reconnection.

\subsubsection{Lack of Tail Flaring}

Two lines of evidence support the inference of a magnetotail lacking substantial flaring during the passage of the ICME. 
First, the topology of the tail current sheet can be inferred from the direction of the field lines. A high amplitude of the $B_Y$ and $B_Z$ components relative to the $B_X$ component imply that the tail is flared and can suggest tail loading. We calculated $|B_Y|/|B_X|$ (and $|B_Z|/|B_X|$) in different regions of the tail (following \citet{Slavin2010}), but no tail loading events were detected. We also calculated an average $|B_Y|/|B_X|$ (and $|B_Z|/|B_X|$) over the tail region for orbit 2577. For orbit 2576 (i.e., prior to the ICME arrival) the average $|B_Y|/|B_X|$ was 0.27 whereas on orbit 2577 the average $|B_Y|/|B_X|$ was 0.13 ($|B_Z|/|B_X|$ was even lower). This results from the dominant and large amplitude $B_X$ on orbit 2577. The large static/thermal pressure in the sheath compressing the tail likely contributed to the observed lack of tail flaring at this time. These results imply lower tail flaring on orbit 2577 than on orbit 2576, indicating that tail loading with magnetic flux was not occurring during orbit 2577.

Second, the shape of the magnetopause boundary in the tail also provides information on the tail geometry. We fit the \citet{Shue1997} magnetopause model to the  magnetopause boundary crossings inbound (high latitude northern hemisphere) and outbound (southern hemisphere) on this orbit to obtain the shape of the magnetopause, and therefore the magnetotail on this orbit (as described in Section 3 and shown in Figure 4). The initial fit yielded a slightly flared ($\alpha = 0.58$; $\alpha=0.5$ indicates no flaring) magnetopause due to the location of the outbound magnetopause crossing in the southern hemisphere. However, tilting of the tail southward due to non-radial ICME flows can cause the outbound magnetopause crossing to be observed farther south than it would be if there were no tail tilting, which can be mistaken for flaring of the magnetotail in an empirical fit. 

We confirmed that the tail was actually tilted southward and not flared by comparing the location of the magnetic equator on this orbit with the average magnetic equator crossing location. Mercury's magnetic equator is observed on average \citep{Johnson2019} at $Z=484\pm 3$~km, whereas on orbit 2577 the northernmost and southernmost crossings of the magnetic equator occurred at $Z=463$~km and $Z=347$~km respectively, both south of the average value. The southward tilting of the magnetotail implies southward tilting of the field lines. To check that this was the case on this orbit, we calculated $\tan(B_Z/B_X)$ in the tail and found a $\sim 5^{\circ}$ downward (i.e., southward) deflection of the field lines, in line with a southward tail tilting.

Further support for tilting of the tail is the fact that there were strong non-radial flows in the ICME. Although we do not have solar wind velocity and density data for this ICME at Mercury, the solar wind velocity at STEREO A shows strong excursions both north and south from radial flow (i.e. away from the ecliptic) during this ICME passage. We determined $\theta_{north-south}$, as described in \citet{Anderson2012} and found it to range between $-9^{\circ}$ and $12^{\circ}$ at STEREO A (see Figure 2), larger than than the average non-radial solar wind flow at Mercury of $4^{\circ}$. As shown in \citet{Anderson2012}, an average non-radial flow at Mercury is able to cause tail tilting. Given that we observed the magnetic equator southward of the average location, we can assume that at the time of the orbit in question MESSENGER was in the portion of the ICME that had a southward directed flow, with an average $\theta_{north-south}$ of $\sim -5^{\circ}$. We thus assumed $\theta_{north-south} = -5^{\circ}$ during this time at Mercury, which is also supported by the observed degree of field line tilting in the tail as mentioned above. We corrected the outbound magnetopause and bow shock crossing points for this extra aberration, by rotating the points clockwise by $\theta_{north-south}$. Fitting the \citet{Shue1997} model to the corrected magnetopause location yields a boundary with a non-flared magnetotail, i.e. a magnetopause with a flaring parameter of $\alpha=0.51$, consistent with the direct observations of extremely low $|B_Y|/|B_X|$ and $|B_Z|/|B_X|$ ratios at this time in the tail.

\subsubsection{Tail Flux and Tail Current}

We calculated the total magnetic flux in the tail from $\Phi_{Tail} = 0.5\pi B_{Tail}R^2_{Tail}$, where $R_{Tail}$, the cross-sectional radius of the tail, is $3800\pm600$~km and was calculated from the location of the midpoint of the outbound magnetopause crossing in the southern hemisphere (after correcting for the southward tilt of the tail). The uncertainty represents the inner and outer limits of the magnetopause layer, outside of which MESSENGER was either entirely in the magnetosheath or in the magnetosphere. $B_{Tail}$ was found to be $114.0\pm5.2$~nT and was calculated by averaging the magnitude of the magnetic field over 2 minutes prior to the outbound magnetopause crossing, with the  uncertainty being the standard deviation of the field over the averaging period. This yields a total magnetic flux in the tail of $2.6\pm0.8$~MWb, consistent with the average value of $2.6\pm0.6$~MWb observed at Mercury \citep{Johnson2012}. 

\begin{figure}
\begin{center}
\includegraphics[scale=0.6]{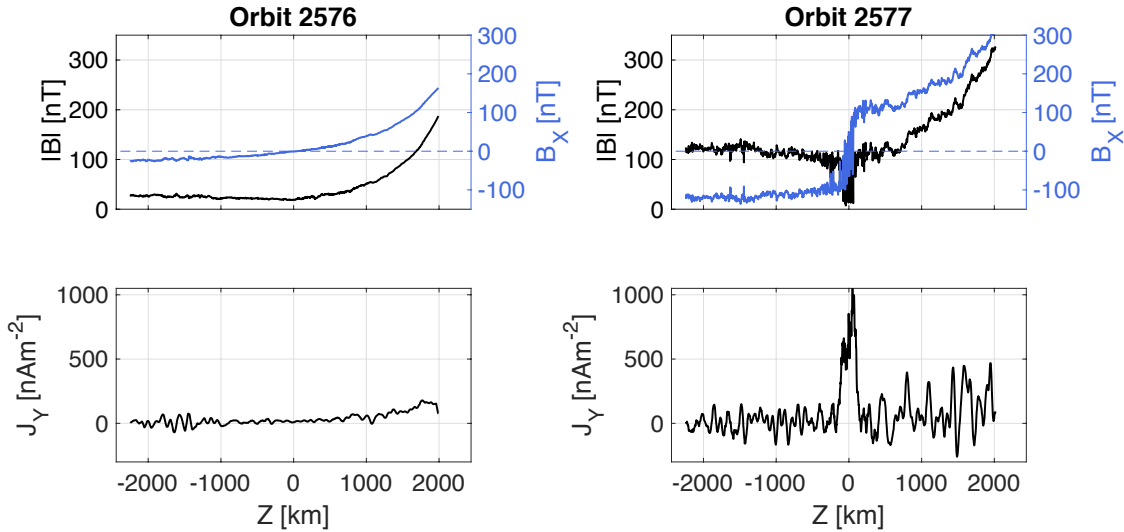} 
\caption{{\bf{Magnetic field data and inferred current densities for orbits 2576 (left column) and 2577 (right column) versus distance in the $Z$ direction.}} The figures are centered on the magnetic equator crossing ($Z$ = 0 km). Top row:  $B_X$ (blue) and $|B|$ (black) in nT. Bottom row: $J_Y = \Delta B_X/ \Delta Z$.} 
\end{center}
\end{figure}

Tail current sheet (TCS) crossings were identified via a clear rotation in the field direction, specifically a rotation in $B_X$ from sunward in the northern lobe to antisunward in the southern lobe, accompanied by a decrease in the total field strength $|B|$ and an increase in the high-frequency variability in the field \citep{Slavin1985}. The magnetic field data from orbit 2576 (Figure 6, top left panel) showed no evidence for a TCS crossing:  $B_X$ changes smoothly along the orbit as expected for a mainly dipolar internal field, and there was no decrease in $|B|$ near the equator crossing.  Assuming that any TCS current flows mainly in the $Y$-direction, the current density, $J_Y$ is proportional to $\Delta B_X/ \Delta Z$. We computed $\Delta B_X/ \Delta Z$ numerically along the orbit, and smoothed the estimate with a 30 second running mean. The resulting $J_Y$ for orbit 2576 confirms no TCS crossing, i.e., on this orbit the magnetic equator crossing was planetward of the TCS. In contrast orbit 2577 (Figure 6, right) had a very different signature. The field magnitude in the tail lobe was $\sim$125~nT, $\sim$4 times that for the previous orbit. A pronounced dip in $|B|$ accompanied a reversal in $B_X$ near the magnetic equator crossing. The inferred current density, $J_Y$ showed a clear peak near $Z$ = 0 km, reaching a maximum value over 1000~nA~m$^{-2}$ (Figure 6, bottom right panel). This confirms that the tail current moved closer to the planet relative to the previous orbit and that the current density was particularly intense, being an order of magnitude higher than during average solar wind at Mercury \citep{AlAsad2018}.

 This observed high current sheet density is also a consequence of the very high ICME \Pram\ that collapsed Mercury's dayside magnetosphere on this orbit. It has been statistically established at Earth \citep{FairfieldJones1996}, and also recently at Mercury \citep{AlAsad2018}, that an increase in \Pram\ increases the tail lobe field and therefore the magnetic pressure in the tail. The increased magnetic pressure in the northern and southern tail lobes causes the current sheet to thin and compresses the total current into a smaller volume thereby increasing the current density. The strong magnetotail current was likely the main source of the increase in the magnetospheric field at closest-approach resulting from the strengthened $B_X$ component.

\section{Discussion and Conclusions}\label{sec:conclusion}

In this paper, using MESSENGER measurements, we analyzed the response of Mercury’s magnetosphere to the highest solar wind ram pressure ever estimated at Mercury. We have confirmed this estimated value of the ram pressure by using  a complementary dataset from STEREO-A which was in longitudinal conjunction with MESSENGER at this time. The magnetosphere exhibited never-before observed characteristics due to the passage of an ICME. These observed characteristics of the magnetosphere include the compression of Mercury's dayside magnetosphere to the surface (Figures 3, 4, and 5), combined with a $\sim40\%$ increase in the spacecraft closest approach magnetic field magnitude, as well as a $\sim300\%$ increase in the magnetotail magnetic field strength (Figure 1) compared to undisturbed conditions.
First, we presented detailed observational evidence which conclusively show the compression of Mercury's dayside magnetosphere on this orbit. Although we used the \citet{Shue1997} model to describe our magnetopause shape, we have clear observations that are not reliant on any specific model that support the inference of the lack of dayside magnetosphere on this orbit. These observations are:
\begin{itemize}
    \item the dayside magnetopause was very close to the surface at high latitude (84$^{\circ}$ N) (Figure 4);
    \item the dayside magnetopause was not crossed earlier in the orbit (at lower latitudes) when MESSENGER was closer to the surface (closest approach occurred at 76.7$^{\circ}$ N) (Figures 1 and 7a top panel);
    \item  complete absence of the dayside dipole field in the magnetosphere (Figure 3); 
    \item a thin dayside magnetopause layer (lasting $<$2 mins) (Figure 1);
    \item dayside bow shock was also very close to the surface at high latitude (Figure 4);
    \item $P_{ram}$ calculated using the proxy method and scaling from STEREO-A both result in a value much higher than needed to collapse Mercury's weak magnetosphere.
\end{itemize}

In terms of the large field enhancements during this orbit, we showed that these are not primarily due to induction in Mercury's iron rich core, but due to the very strong (order of magnitude higher than normal) magnetotail current density at this time. The highly increased magnetotail fields, caused by the extremely high ICME $P_{ram}$, are likely the largest additions to the planetary dipole field providing the $\sim40\%$ increased peak field observed by MESSENGER on this orbit.

Furthermore, we showed that the lack of tail flaring as well as an average tail flux during this orbit support the fact that no significant dayside reconnection occurred on this orbit. Therefore, the absence of the dayside magnetosphere was not due to reconnection (as previously thought to be the main mechanism for such magnetospheric collapse), but due to compression from the increased pressure associated with the ICME at this time.

\begin{figure}
\includegraphics[width=\linewidth]{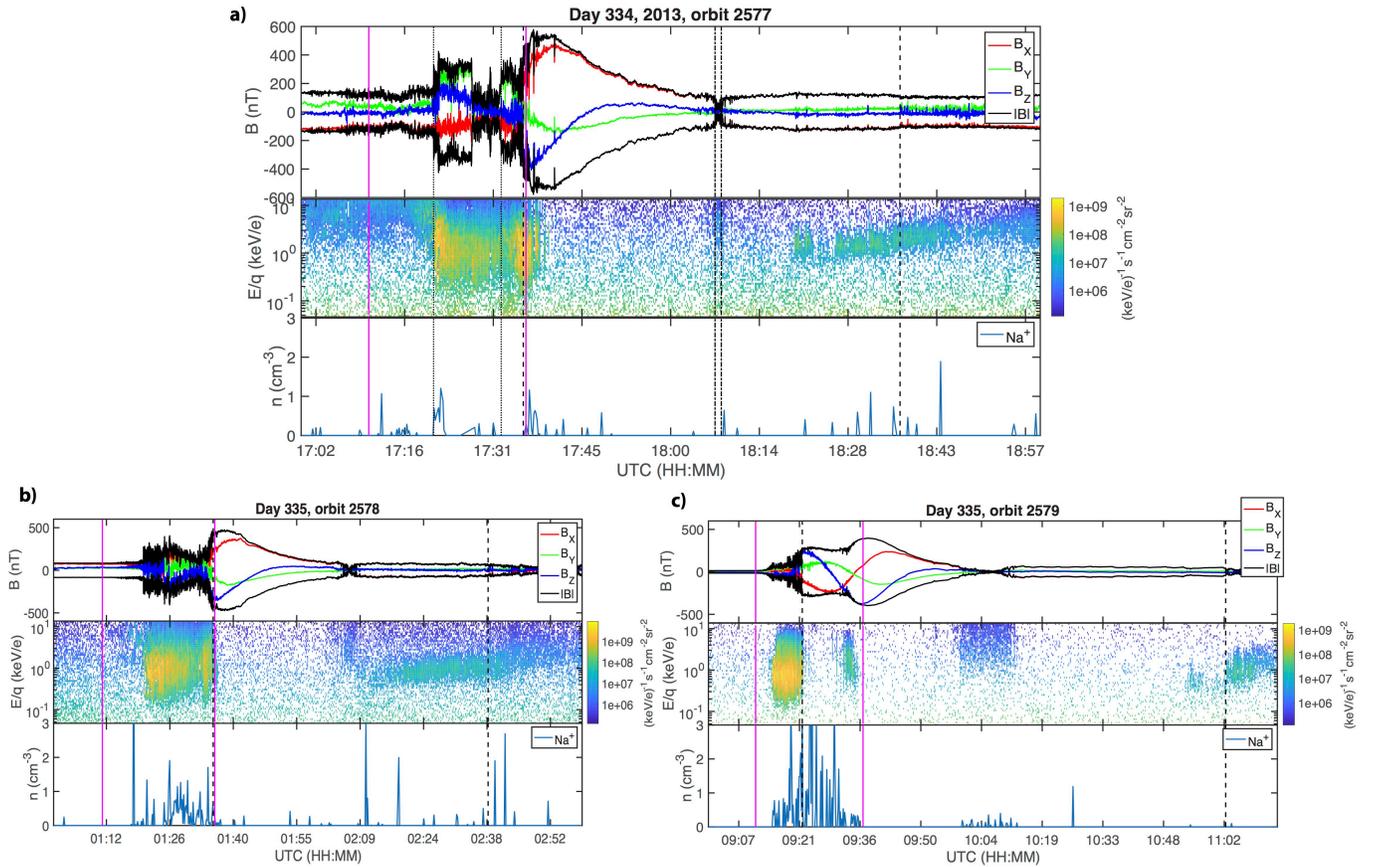} 
\caption{{\bf{Proton and Na$^+$ observations by FIPS during three consecutive orbits.}} (a) Magnetic field data versus time, proton differential energy flux versus energy per charge (E/q) as a function of time, and Na$^+$ density versus time for orbit 2577. The first magenta line marks the planetary equator, the second magenta line is near the North pole ($\sim$84$^{\circ}$), the dotted black lines mark the inner and outer limit of the inbound bow shock boundary, the dashed black lines mark the magnetopause boundary, and the dot dashed lines mark the beginning and end of the magnetic equator. In the magnetotail, plasma was observed both near the magnetic equator crossing (consistent with the plasma sheet) and in the southern tail lobe prior to the outbound magnetopause crossing. There were no Na$^+$ density increases observed near equatorial latitudes on the dayside. (b) Same as in a) but for orbit 2578. Large density increases were observed near equatorial latitudes on the dayside.(c) Same as a) but for orbit 2579. Large density increases were observed near equatorial latitudes on the dayside. } 
\end{figure}

Recent observations of the exospheric sodium emission pattern at Mercury detected diffused equatorial sodium at times of ICME passage in general, in contrast to the more commonly observed two-peak pattern at high latitudes in both hemispheres \citep{Orsini2018}. The major driver of sodium surface release is particle precipitation, thus it is expected that during average conditions most of the sodium emission is in the cusp regions where the magnetic field is weakest and particle precipitation is common \citep{Winslow2012,Winslow2014,Raines2014,Poh2016}.
However, compression of Mercury's dayside magnetosphere to the surface allows ICME plasma to directly impact the dayside at low latitudes, thereby sputtering sodium into the exosphere. Our observation here of such compression provides evidence for a mechanism of sodium-generation at equatorial latitudes.

We checked whether the FIPS Na$^+$ observations for this ICME are in line with the ground-based sodium observations during ICMEs by \citet{Orsini2018}. On orbit 2577, FIPS observed no dayside equatorial Na$^+$ (Figure 7a), which is consistent with the collapse of the magnetosphere and the accompanying dominance by ICME particles on this orbit. However, the following two MESSENGER orbits (Figures 7b and c) showed increased densities over the pre-ICME levels on most of the dayside, indicating Na$^+$ generation from sputtering by the high density ICME. The increased Na$^+$ densities are observed extending from low to high latitude, in line with the results of \citet{Orsini2018}, confirming increased exospheric generation during this extreme ICME event.

Additionally, the disappearance of Mercury's dayside magnetosphere may have important implications for exoplanetary systems. M dwarfs, the most common type of star \citep{Henry2006}, can have high occurrence rates of stellar flares \citep{Odert2017} and are known to typically host exoplanets \citep{DressingCharbonneau2013}. It is unknown whether the flare-to-CME ratio of M dwarfs is similar to that of the Sun \citep{Odert2017}. However, because the habitable zone of M dwarfs is $\sim$10 times closer to the star than in our solar system, exoplanets in these zones are susceptible to possibly very frequent magnetic eruptive events. ICMEs at close proximity to the host star are likely to carry higher \Pram\ than that observed in this study, due to the increased ICME speed and density close to the star. We thus suggest that a habitable exoplanet around an M dwarf might require an intrinsic planetary magnetic field substantially stronger than that of Mercury to not lose its atmosphere over geologic timescales. It has also been suggested that tidally locked planets \citep{Kite2011} may have inherently weaker magnetic moments \citep{Khodachenko2007}, so that for close-in exoplanets stellar-flare activity and associated ICME interactions may significantly hinder the emergence and development of life. Based on our results, we thus recommend that modeling of exoplanet atmospheric loss due to space weather \citep{Dong2017a,Dong2017b,GarciaSage2017} take into account the possibility of magnetospheric collapse events. 

Finally, the two spacecraft BepiColombo mission \citep{BENKHOFF2010} arriving at Mercury in 2025, during the next solar maximum, will be well positioned to observe such magnetospheric collapse events in more detail. The presence of the Mercury Magnetospheric Orbiter (MMO) spacecraft combined with the Mercury Planetary Orbiter (MPO) spacecraft will greatly simplify the analysis of future such measurements, as the MMO will act as an upstream solar wind monitor approximately $80\%$ of the time (i.e., when it is projected to be outside of Mercury's magnetosphere). In the 1 year nominal and 1 year extended mission of BepiColombo, we can expect a total of $\sim35$ ICME events based on the number of ICMEs detected by MESSENGER \citep{Winslow2015,Winslow2017} in the same part of the solar cycle. Out of these, $\sim10$ events, or $\sim30\%$ \citep{Winslow2017}, will likely collapse at least part of Mercury's dayside magnetosphere to the surface. This additional dual spacecraft dataset of such events will greatly enhance our understanding of the physical processes at play during these extreme conditions affecting the solar system's innermost planet.

\acknowledgments
Support for this work was provided by NASA grant NNX15AW31G. R. M. W. acknowledges support from NASA grants NNX15AW31G and 80NSSC19K0914, and NSF grant AGS1622352. N. L. acknowledges support from NASA grants NNX15AB87G and NNX13AP52G. C. J. F. was partially supported by the NASA STEREO Quadrature grant. C. L. J., M. A. A. and L P. acknowledge support from the Natural Sciences and Engineering Research Council of Canada. We would like to thank the STEREO PLASTIC PI for the STEREO He/H data provided for this study.

 All the data analyzed in this study are publicly available. MESSENGER data are available on the Planetary Data System ({\url{https://pds.jpl.nasa.gov}}), while STEREO data are available on the Space Physics Data Facility's Coordinated Data Analysis Web ({\url{https://cdaweb.sci.gsfc.nasa.gov}}).

\bibliography{Refs}{}
\bibliographystyle{aasjournal}



\end{document}